\documentclass[twocolumn,10pt]{article}
\usepackage{graphics}
\usepackage{epsfig}
\usepackage{cite}

\makeatletter \renewcommand\@biblabel[1]{#1.}\makeatletter

\makeatletter
\def\@eqnnum{{\normalfont \bf \normalcolor [\theequation]}}
\makeatother

\begin{document} 
%
\title{Ion channel gating: a first passage time analysis of the Kramers type}
\author{Igor Goychuk and Peter H\"anggi\\Institute of Physics, 
University of Augsburg,\\
Universit\"atsstr. 1, D-86135 Augsburg, Germany}
 \maketitle
{\bf 
\noindent   
ABSTRACT. The opening rate  of voltage-gated potassium ion channels
exhibits a characteristic, knee-like turnover where the common
 exponential voltage-dependence changes  suddenly
into a linear one.
An explanation of this puzzling crossover is put forward in terms of 
a stochastic first passage time analysis. 
The theory predicts that the exponential voltage-dependence
correlates with the
exponential distribution of closed
residence times. This feature occurs at large negative voltages 
when the channel is
predominantly closed. 
In contrast, the linear part of voltage-dependence
emerges together with a non-exponential distribution of closed dwelling 
times with increasing voltage, yielding a large opening rate. Depending on the 
parameter set,
the closed-time distribution  displays a power law behavior
which extends  over  several decades. }

 
\section*{Introduction}

Voltage-dependent ion channels of biological membranes are formed
by pore-like single proteins which poke through the cell membrane. 
They   provide the conducting pathways for the ions of specific sorts 
\cite{hille,neher}. Such 
potassium ${\rm K}^{+}$ and sodium ${\rm Na}^{+}$ channels 
participate in many important processes 
occurring in living
cells. For example, these are crucial for the phenomenon 
of neural excitability \cite{HodHux52}. 

Two features are important for the biological function of 
these naturally occurring
nanotubes. First, they either are dwelling in open conformations, allowing
for the ion flow to pass through, or are resting  in closed, non-conducting
conformations. Between these two conformation types  the ion channel undergoes 
spontaneous, temperature driven transitions -- the so-called gating 
dynamics -- which can be characterized
by the residence time distributions of open, $f_o(t)$, and closed, $f_c(t)$,
states, respectively. The mean open and closed residence times,
$\langle T_{o(c)} \rangle:=\int_{0}^{\infty}tf_{o(c)}(t)dt$ 
are prominent
quantifiers of the gating dynamics. In particular, they determine
the mean opening (closing) rates $k_{o(c)}:=\langle T_{c(o)} \rangle^{-1}$. 
The second important feature refers to the fact that
 the gating dynamics is voltage-dependent. This 
provides a mechanism for a mutual coupling among otherwise independent ion 
channels. This very mechanism is realized 
through the common membrane potential. Both ingredients are central for
the seminal model of neuronal activity put forward  by Hodgkin and Huxley
in 1952 \cite{HodHux52}.

The dichotomous character of gating transitions yields a bistable
 dynamics of the Kramers type \cite{HTB90}. Therefore,  {\it a 
 priori\/} one expects that both, the opening and the closing gating rates 
 will expose
an exponential, Arrhenius-like dependence on voltage and temperature.
Indeed, the closing rate of many ${\rm K}^{+}$ channels follows such 
a pattern; in clear contrast, however, the opening rate usually does not.
To 
explain the experimental voltage-dependence of the relaxation time
of the potassium current
for a giant squid axon  Hodgkin and Huxley\cite{HodHux52} 
postulated that the 
gating behavior of a potassium channel is determined by four independent
voltage-sensitive gates, each of which undergoes a two-state Markov dynamics
with  a form \cite{HodHux52,mainen} 
\begin{equation}\label{ko}
k_o(V) = \frac{a_c(V-V_c)}{1-\exp[-b_c(V-V_c)]}
\end{equation}
for the opening rate, which is commonly used in neurophysiology.
In Eq. {\bf\ref{ko}}, $a_c,b_c, V_c$ are some experimental 
parameters. Notwithstanding that in their work \cite{HodHux52}
this kind of dependence has been used for a single gate,
the opening rate of the {\it whole} ${\rm K}^{+}$ channel can also be fitted
by Eq. {\bf\ref{ko}}, see e.g. in \cite{marom}. 
The same 
modeling for a whole channel is used also for 
dendritic ${\rm K}^{+}$ channels in
neocortical pyramidal neurons \cite{mainen}. 

Note that in Eq. {\bf\ref{ko}}
the voltage-dependence of the opening rate changes in a knee-like 
manner from an  exponential behavior into a linear one, cf. Fig. 1. 
\begin{figure}
\begin{center}
\epsfig{file=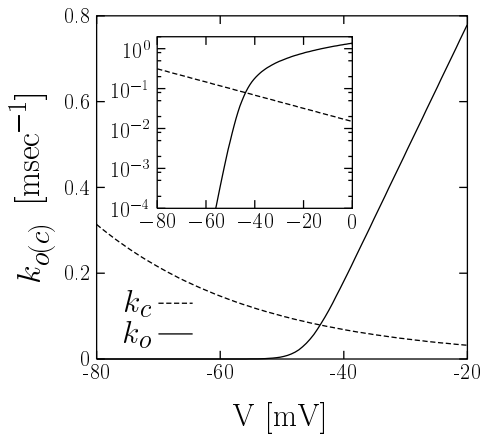,width=0.4\textwidth}
\end{center}
\caption{Dependence of opening ($k_o$) and closing ($k_c$) rates 
on voltage for a {\it Shaker IR} ${\rm K}^{+}$ channel from
\cite{marom} at $T=18^o{\rm C}
$. The opening rate is described by Eq. (1) with the following
parameters \cite{marom}: 
$a_c=0.03 \;{\rm msec/mV}$, $b_c=0.8\; {\rm mV^{-1}}$
and $V_c=-46\; {\rm mV}$. The closing rate is given by $k_c=0.015\exp(-0.038V)$
${\rm msec}^{-1}$ ($V$ in mV)
\cite{marom,GH00}. Inset shows the same dependencies in the
semi-logarithmic scale. }
\label{Fig2}
\end{figure}
This typical, experimentally observed    behavior of
delayed rectifier ${\rm K}^{+}$ channels 
presently lacks an explanation in physical terms.
A  qualitative explanation  of this gating
dynamics has briefly been mentioned in  recent work \cite{sigg}. However, 
a definite analysis
leading to the functional form in Eq. {\bf\ref{ko}} is not available.
A {\it first} main objective of the present work is to fill this gap, 
and, moreover, to 
 provide
additional insight into the
 voltage behavior of Eq. {\bf\ref{ko}} within an {\it exactly} solvable
stochastic Fokker-Planck-Kramers model.

The ion current recordings made on the level of {\it single}
ion channels \cite{neher}
reveal yet another unresolved, interesting aspect of the gating dynamics. 
Namely, the
 distribution of closed residence times of many channels is {\it not} 
exponential.  
In particular, it has been shown
by Liebovitch {\it et al.} \cite{lieb87} that the closed residence time
distribution $f_c(t)$ in a rabbit corneal endothelium  channel
can be reasonably fitted by a stretched exponential with
only two parameters. This result initiated the construction of the
so-called fractal model of  ion channel gating \cite{lieb87,dewey}. 
Other channels, e.g., ${\rm K}^{+}$ channels in neuroblastoma x
glioma (NG 108-15) cells  exhibit a power-law scaling 
behavior as well, i.e. $f_c(t)\propto t^{-\alpha}$ with $\alpha=3/2$ 
\cite{sansom}.
To explain this type of fractal-like behavior Millhauser {\it et al.} 
\cite{millhauser} proposed a one-dimensional
diffusion model. Similar power laws with $\alpha>3/2$  have also been 
reported \cite{add} and several variations of diffusion theory
have been introduced to explain the gating behavior
of different channels \cite{lauger,condat,levitt}. 

The observed non-exponential 
behavior can be fitted by a finite sum of exponentials; consequently,
it can  alternatively be explained with a corresponding  discrete Markovian 
scheme \cite{sansom}. 
These discrete Markovian models
have proven  their usefulness in many cases 
\cite{colquinon}.  Nevertheless, such an approach presents a fitting procedure; 
as such it  is intimately connected
with the danger of a  proliferation of parameters. In particular,
 kinetic schemes containing as many as 14 structurally
unidentified closed substates have been 
proposed \cite{fedida}. 

\begin{figure}
\begin{center}
\epsfig{file=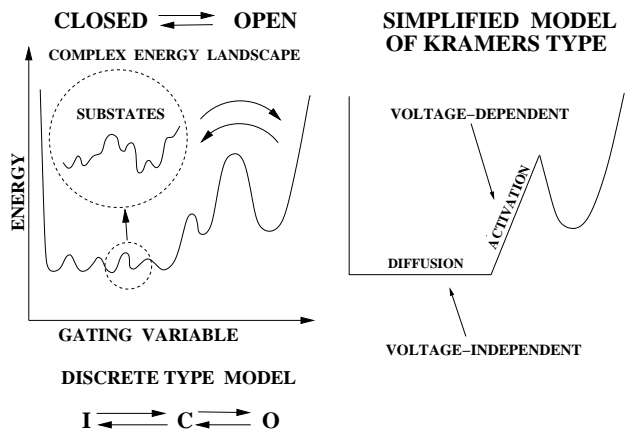,width=0.5\textwidth}
\end{center}
\caption{Gating dynamics as an activated diffusion on a complex
free energy landscape. Two global minima correspond to open and closed
macroconformations. One assumes a large number of quasi-degenerated
(within $k_BT$) and voltage-independent closed substates separated from
the open conformation by a voltage-dependent potential barrier. This
idea is sketched by a simplified model of the Fokker-Planck-Kramers type,
 and by a discrete
 model with open (O), closed (C) and inactivated (I) states. }
\label{Fig1}
\end{figure}

An important lesson to be learned from the detailed studies of
a simple protein -- myoglobin -- by Frauenfelder {\it et al.}
\cite{frauenfelder}
is that proteins
exist in a huge number of quasi-degenerated, microscopic 
substates, corresponding to a single macroscopic conformation, 
cf. Fig.~2. It is thus conceivable that at room 
temperatures
the ion channel dwells in a huge number of almost degenerated 
(within $k_BT$) conformational substates. Both the fractal and diffusion 
models of the ion channel gating have
been inspired by this crucial property of proteins.
We conjecture that the ultimate 
theory of the ion channel
gating must take this property into account.  This program requires 
a compromise between  Markovian discrete state
models and a continuous diffusion model. This can be achieved
 by a Kramers type theory \cite{sigg, HTB90}. The discrete Markov models can
then be considered as a limiting case of more general Kramers type 
approach \cite{HTB90}.

\section*{Theoretical modeling}

The complex structure of the multi-dimensional conformational
space of proteins implies an intricate kinetics despite an apparently
simple bistability that is observed
\cite{frauenfelder}. Two popular theoretical approaches have
been developed to cope with this complexity. A first one
uses a simple bistable dynamics as a basis. To model the
complexity of the observed kinetics this dynamics is amended
by using an additional  stochastic time-dependence of the 
energy profile, or kinetic constants. Such
an approach is nowadays commonly known under the label of ``fluctuating
barriers'' \cite{fluct}. Alternatively, one can attempt to 
model the complexity of the energy profile 
itself in the simplest possible way.
Our strategy is to find such a minimal 
model of the second kind which does allow for a rigorous analysis
and does reproduce some nontrivial 
features of the gating dynamics.

Let us assume that the conformational stochastic dynamics between
the open and closed states  can be described in terms of a 
one-dimensional reaction coordinate dynamics $x(t)$  in a conformational 
potential
$U(x)$, Figs. 2,3. Since the 
distribution of open 
residence time intervals assumes typically a single-exponential \cite{hille},
in the following we rather shall focus on the behavior of the closed residence 
time 
intervals. In order to evaluate the distribution
of closed residence time intervals it suffices  
to restrict our analysis to
the subspace of closed states by putting an absorbing
boundary at the interface, $x=b$, between the closed and open
conformations, see Fig. 3.
We next assume
that the gating dynamics is governed by two gates: an inactivation
gate and an activation gate. The inactivation gate corresponds to 
the manifold of {\it voltage-independent} closed substates. 
It is associated with
the flat part, $-L<x<0$,  of the potential $U(x)$ in Fig. 3.
In this respect, our modeling  resembles 
that in \cite{sigg97}. 
The mechanism of inactivation in potassium
channels is quite sophisticated and presently not totally
established \cite{hille}. It is well known that inactivation can occur
on quite different time scales \cite{hille}. The
role of a fast inactivation gate in  
{\it Shaker} ${\rm K}^{+}$ channels 
is taken over by the channel's 
extended N-terminus which is capable to plug the channel's pore from the 
cytosol part while
diffusing towards the pore center \cite{mckinnon}. The slow
inactivation apparently is due to a conformational
narrowing of the channel pore in the region of selectivity
filter \cite{hille}.  
In both cases, no net gating charge translocation occurs
and the inactivation process does not depend on voltage.
When the inactivating
plug is outside of the pore, or the selectivity filter is open 
($x>0$ in Fig. 3) the channel can
 open only {\it if} the activation barrier is overcome. 

The dynamics of the activation gate  occurs on
the linear part of the ramp of the potential $U(x)$; 
i.e. on $0<x<b$ in Fig. 3, like in  \cite{levitt}. 
Note that for $0<x<b$, the inactivating plug diffuses
outside of the channel's pore and the selectivity filter is open. 
 During the activation step
a gating charge $q$ moves across the membrane, this feature renders 
the overall 
gating dynamics voltage-dependent. The channel opens
when the reaction coordinate reaches the location
$x=b$ in Fig. 3. This fact is accounted for by putting   
an absorbing boundary condition at $x=b$. Moreover, the channel
closes immediately when the inactivation gate closes ($x\leq 0$),
or when the activation gate closes.
To account for this behavior in  extracting  the closed residence time 
distribution 
we assume that the channel is reset into the
state $x=0$ after each closure (see below).

The diffusional motion of the inactivated gate 
is restricted in conformational space. 
We characterize this fact
by the introduction of a conformational diffusion length $L$ (Fig. 3) and 
the diffusion constant $D\sim k_BT$
that are combined into a single parameter -- the conformational 
diffusion time
\begin{eqnarray} \label{tauD}
\tau_D=L^2/D \;.
\end{eqnarray}
This quantity constitutes an essential parameter for the theory. 
\begin{figure}
\begin{center}
\epsfig{file=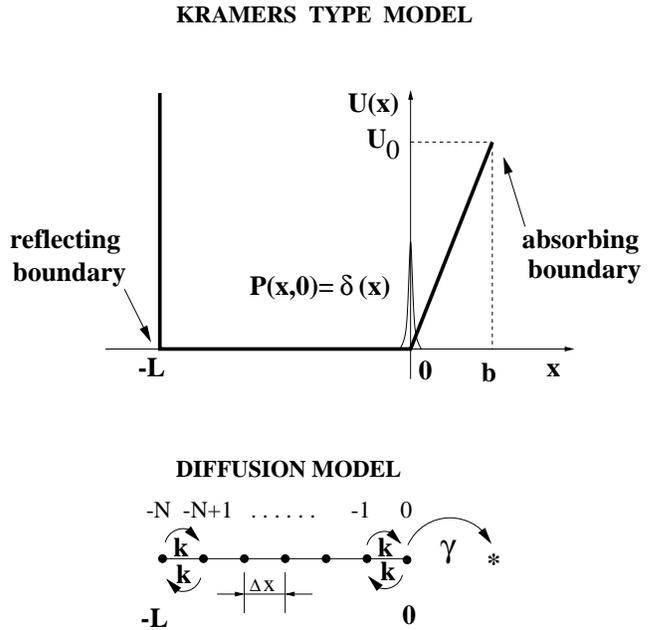,width=0.5\textwidth}
\end{center}
\caption{Studied model and its diffusion
counterpart}
\label{Fig3}
\end{figure}
We 
assume that the activation barrier height $U_0$ is linearly 
proportional to the 
voltage bias $V$ \cite{levitt},  i.e. in terms of the gating charge 
$q$ we have 
\begin{equation}\label{U0}
U_0=-q(V-V_c).
\end{equation}
Moreover,  $U_0$ is positive for negative voltages, i.e. for 
$V<V_c$, vanishes at $V=V_c$, and becomes negative for $V>V_c$.
Thus, for $V>V_c$ the channel ``slips'' in its open state, rather
than overcomes a barrier. 
In addition, 
the fraction $\xi$ of the voltage-dependent substates 
in the whole manifold of the closed states should be
very small, implying that
\begin{equation}\label{xi}
\xi=b/L\ll 1.
\end{equation} 

 
{\bf Analytical solution}. 
The corresponding Fokker-Planck equation for the probability density 
of closed states $P(x,t)$ reads
 \begin{eqnarray}\label{kramers}
\frac{\partial P(x,t)}{\partial t}=D\frac{\partial}{\partial x}
\left(\frac{\partial}{\partial x}+\beta\frac{\partial U(x)}{\partial x}
\right) P(x,t),
\end{eqnarray}
where $\beta=1/(k_BT)$. In order to find the distribution of closed 
residence times
$f_c(t)$, we  solve Eq. {\bf\ref{kramers}}
with the initial condition $P(x,0)=\delta(x)$, in combination with 
a reflecting boundary 
condition 
$\frac{dP(x,t)}{dx}|_{x=-L}=0$,
and an absorbing boundary condition, $P(x,t)|_{x=b}=0$ \cite{HTB90}.
The closed residence time distribution then follows as
\begin{equation}\label{def1}
f_c(t)=-\frac{d\Phi_c(t)}{dt},
\end{equation}
where $\Phi_c(t)=\int_{-L}^{b}P(x,t)dx$ 
is the survival probability  of  the closed state.

By use of the standard Laplace transform method 
we arrive at the following  
{\it exact} solution:
\begin{equation}\label{result4}
\tilde f_c(s)=\frac{A(s)}{B(s)},
\end{equation}
where 
\begin{eqnarray}\label{result5}
A(s) &= & 
\exp(-\beta U_0/2)\sqrt{\beta^2 U_0^2+4\xi^2\tau_D s} \\ \label{result6}
B(s)& = &\sqrt{\beta^2 U_0^2+4\xi^2\tau_D s}\\ \nonumber
&\times& \cosh\Big( \sqrt{
\beta^2 U_0^2+4\xi^2\tau_D s}/2\Big)\\ \nonumber
&+&\Big(2\xi\sqrt{\tau_D s}
\tanh\sqrt{\tau_D s}-\beta U_0 \Big) \\ \nonumber
&\times& \sinh\Big( \sqrt{
\beta^2 U_0^2+4\xi^2\tau_D s}/2\Big). 
\end{eqnarray}
 The explicit result in {\bf\ref{result4}-\ref{result6}} allows one to find 
all moments of
the closed residence time distribution. In particular, the
mean closed residence time $\langle T_c\rangle=\lim_{s\rightarrow 0}
[1-\tilde f_c(s)]/s $ reads 
\begin{eqnarray}\label{result1}
\langle T_c\rangle=\tau_D\xi
\frac{\beta U_0(e^{\beta U_0}-1-\xi)+\xi(e^{\beta U_0}-1)}{\beta^2 U_0^2}.
\end{eqnarray}
This very same result {\bf\ref{result1}} can be obtained alternatively 
if we invoke the
well-known relation for the mean first-passage time 
$\langle T_c \rangle=\frac{1}{D}\int_{0}^{b}dx e^{\beta U(x)}
\int_{-L}^{x}dy e^{-\beta U(y)}$ \cite{HTB90}. 
This alternative scheme provides a successful validity check for our 
analytical solution in
{\bf\ref{result4}-\ref{result6}}.

{\bf Elucidation of the voltage dependence in Eq. {\bf\ref{ko}}}. 
Upon observing the
 condition {\bf\ref{xi}}  Eq. {\bf\ref{result1}} by use of
 {\bf\ref{U0}} reads in leading order of $\xi$ 
\begin{equation}\label{result7}
k_o=\frac{1}{\langle T_c \rangle}
\approx\frac{\beta q}{\xi\tau_D}\frac{V-V_c}
{1-\exp[-\beta q (V-V_c)]}.
\end{equation}
With the parameter identifications 
\begin{equation}\label{bc}
b_c=\frac{q}{k_BT}
\end{equation}
 and 
\begin{equation}\label{ac}
 a_c=\frac{q}{\xi\tau_Dk_BT}
\end{equation} 
the result in {\bf\ref{result7}} precisely coincides with 
Eq. {\bf\ref{ko}}. The fact that our novel approach yields 
the puzzling
voltage dependence in Eq. {\bf\ref{ko}} constitutes a first prime result
of this work. 

Let us next estimate the model parameters for a {\it Shaker IR}
 ${\rm K}^{+}$ channel from Ref. 
\cite{marom}. In \cite{marom}, 
the voltage-dependence of $k_o(V)$ at $T=18 \;{\rm ^oC}$ has been parameterized
by Eq. {\bf\ref{ko}} with the parameters given in the caption 
of Fig. 1. Then, from 
Eq. {\bf\ref{bc}} 
the gating charge can be
estimated as $q\approx 20e$ ($e$ is the positive valued,  elementary charge). 
As to
the diffusion time $\tau_D$, we
speculate that it corresponds to the time scale of inactivation; the 
latter is in the range of seconds and larger \cite{marom}. Therefore, 
we  use 
$\tau_D=1\;{\rm sec}$ as a lower bound for our estimate. 
The fraction of voltage-dependent states $\xi$ is then extracted from
Eq. {\bf\ref{ac}} to yield, $\xi\approx 0.0267$. This value, indeed, is 
rather small
and thus proves our finding  in   
Eq. {\bf\ref{result7}} to be  consistent.

{\bf Analysis for the closed residence time distribution}. The exact
results in Eqs. {\bf\ref{result4}-\ref{result6}} appear rather 
entangled. To extract the behavior in real time one needs to invert 
the  Laplace transform numerically. With $\xi << 1 \/$, however,
Eqs. {\bf\ref{result4}-\ref{result6}} are  formally reduced to 
\begin{eqnarray}\label{result2}
\tilde f_c(s)=\frac{1}{1+(k_o\tau_D)^{-1}\sqrt{\tau_D s}
\tanh\sqrt{\tau_D s}}.
\end{eqnarray}
This prominent leading order result  can be inverted {\it analytically} 
in terms of an infinite sum of exponentials, yielding:
\begin{eqnarray}\label{distribution}
f_c(t)=\sum_{n=1}^{\infty}c_n\lambda_n\exp(-\lambda_n t),
\end{eqnarray}
where the rate constants $0<\lambda_1<\lambda_2<...$ are solutions of the 
transcendental equation 
\begin{eqnarray}\label{eigen}
\tan \sqrt{\lambda_n\tau_D}=\frac{k_o\tau_D}{\sqrt{\lambda_n\tau_D}}
\end{eqnarray}
and the expansion coefficients $c_n$, respectively,  are given by
\begin{eqnarray}\label{coeff}
c_n=\frac{2}{1+k_o\tau_D+\lambda_n/k_o}.
\end{eqnarray}
Note from Eq. {\bf\ref{def1}} that the set $c_n$ is normalized 
to unity, i.e.  $\sum_{n=1}^{\infty}c_n=1$.

The analytical approximation, Eqs. {\bf\ref{distribution}-\ref{coeff}},
is compared in Fig. 4 with the precise numerical inversion of the exact 
Laplace transform
in Eqs. {\bf\ref{result4}-\ref{result6}}. The numerical inversion has been
performed with the 
Stehfest algorithm 
\cite{stehfest}. As can be deduced from
 Fig. (4), for $t> 10$ msec the agreement is very good indeed.  
A serious  discrepancy occurs only in the range 
$0.01\;{\rm msec}<t<0.1\;{\rm msec}$
 which lies outside the range of the
 patch clamp experiments ($t>0.1\;{\rm msec}$). Moreover, the agreement
 is improving  with  increasing $\tau_D$ (not shown).

\begin{figure}
\begin{center}
\epsfig{file=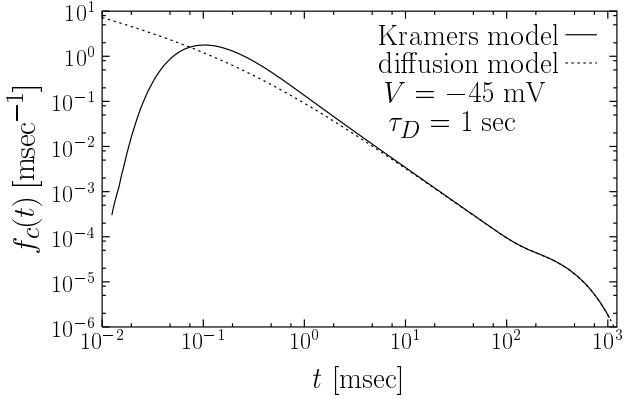,width=0.5\textwidth}
\end{center}
\caption{Closed residence time distribution for a diffusion-limited case.
The exact numerical result (full line) is compared with the analytical
approximation in Eqs. (15)--(17) (broken line). The latter one 
coincides with the exact
solution of the diffusion model by Millhauser {\it et al.} in the scaling
limit.  }
\label{Fig4}
\end{figure}

{\bf Origin of the power law distribution}.
The features displayed by the closed residence time distribution $f_c(t)$
depend sensitively on the applied voltage $V$. When $V>V_c$, e.g. 
$V=-45\; {\rm mV}$\/, as in Fig. 4, the
activation barrier towards the channel opening disappears and
the opening dynamics becomes diffusion-limited. In this case,
the diffusion time $\tau_D=1\;{\rm sec}$ largely exceeds 
the  mean closed residence time $\langle T_c\rangle\approx 18.4\;
{\rm msec}$. Put differently,
$ \tau_D\gg \langle T_c\rangle$ and the closed residence time distribution
exhibits an intricate behavior with three distinct regions, see in Fig. 4.
Most importantly, for the
intermediate time scale 
\begin{equation}\label{inter}
\langle T_c\rangle^2/\tau_D\ll t\ll \tau_D
\end{equation}
we find from Eq. {\bf\ref{result2}} (by considering 
the limit $\tau_D\to\infty$) 
that the closed residence time distribution obeys a
power law; reading
\begin{equation}\label{approx1}
f_c(t)\approx \frac{1}{2(\pi\tau_D)^{1/2}k_ot^{3/2}}.
\end{equation} 
This type of behavior is clearly detectable in Fig. 4 where it covers
about two decades of time. 
As follows from Eq. {\bf\ref{inter}}, an increase of $\tau_D$ by {\it one} order
of magnitude (while keeping $\langle T_c\rangle$ fixed) 
 extends the power law region by {\it two}
orders of magnitude. This conclusion is fully confirmed by our numerics
(not shown).  This power law dependence, which extends
 over four orders of magnitude,
has been seen experimentally for a ${\rm K}^{+}$
channel in NG 108-15 cells \cite{sansom}. On the contrary, for 
channels, where $\tau_D$ is smaller, the power law region {\bf\ref{inter}} 
shrinks and eventually disappears, whereas the mean
opening rate defined via Eq. {\bf\ref{result1}} still exhibits a steep
dependence on the voltage. Thus, our model is capable to describe for 
different channels both, 
 the emergence of power law as well as its absence.

On the time scale $t\geq \tau_D$ the discussed 
power law distribution crosses over into the
exponential tail; the latter is 
fully described by the smallest exponent $\lambda_1$
in Eq.{\bf\ref{distribution}}, i.e., by  
\begin{equation}\label{approx2}
f_c(t)\approx c_1\lambda_1\exp[-\lambda_1t].
\end{equation}
This feature is  clearly manifest in Fig. 4. The transition
towards the exponential tail in the closed residence time-interval distribution 
can be
used to estimate the diffusion time $\tau_D$ on pure {\it 
experimental} grounds!

Finally, let us consider the opposite limit, $\tau_D\ll\langle T_c\rangle$,
for $V\ll V_c$.
For the considered set of parameters this occurs, e.g.,  for $V=-55 \;{\rm mV}$
when the channel is predominantly closed. Then,
the diffusion step in the opening becomes negligible and in the 
experimentally
relevant range of closed residence times, defined by $\langle T_c\rangle$, 
the corresponding distribution can
be approximated by a single exponential {\bf\ref{approx2}}.  A 
perturbation
theory in Eq. {\bf\ref{eigen}} yields $\lambda_1\approx k_o(1-(k_o\tau_D)/3)$.
For the used parameters we have $\lambda_1\approx 0.96 k_o$ and,
 from Eq. {\bf\ref{coeff}}, $c_1\approx 0.95$. This result is in a
 perfect agreement with the precise numerics obtained from Eqs. 
 {\bf\ref{result4}-\ref{result6}}.
Thus, the distribution of closed residence times is single-exponential to a 
very good
degree.  Consequently, one and the same
channel can exhibit both, an exponential and a power-law distribution
of closed residence times, as a function of the applied
 transmembrane voltage. With an increase  of $\tau_D$ the voltage
range of the exponential behavior shifts towards more negative 
voltages, $V<V_c$, 
 and {\it vice versa}.

{\bf Reduction to a diffusion model}.
Let us relate our model to that introduced
previously  by Millhauser {\it et al.}
\cite{millhauser}. The latter one is  depicted with the lower part in Fig. 3.  
It  
assumes a discrete number $N$ of closed substates
with the same energy. The gating particle jumps with the equal forward and 
backward rates $k$ between the adjacent
states which are occupied with probabilities $p_n(t)$. 
At the right edge of the chain of closed
states the ion channel undergoes transition into the open state
with the voltage-dependent rate constant $\gamma$. To calculate the
closed residence time distribution $f_c(t)$ one assumes $p_0(0)=1$, $p_{n\neq 
0}(0)=0$
and $d\Phi_c(t)/dt=-\gamma p_0(t)$, where $\Phi_c(t)=\sum_{n=0}^{n=-N}p_n(t)$
is the survival probability \cite{millhauser,condat}.

We consider the continuous diffusion variant
of this model \cite{nadler} in a scaling limit: we put
$\Delta x \to 0, k\to \infty, \gamma\to \infty, N\to\infty$ 
 keeping  the diffusion
length $L=N\Delta x$, the diffusion constant 
$D=k(\Delta x)^2$, and the constant $k_o=\gamma/N$ all finite.
 The latter one has the meaning
of mean opening rate, see below. Note that in clear contrast with our 
approach, 
the rate parameter $k_o$ in the diffusion model is of pure phenomenological 
origin. The problem of finding
the closed residence time distribution is reduced to solving the diffusion
equation 
\begin{eqnarray}\label{difeq}
\frac{\partial P(x,t)}{\partial t}=D\frac{\partial^2 P(x,t)}{\partial x^2}
\end{eqnarray}
with the initial condition $P(x,0)=\delta(x-0_{-})$, the
reflecting boundary condition $\frac{\partial P(x,t)}{\partial x}|_{x=-L}=0$
and the radiation boundary condition \cite{bezrukov} 
\begin{eqnarray}\label{reactive}
\frac{\partial P(x,t)}{\partial x}|_{x=0}=-\frac{Lk_o}{D} P(0,t).
\end{eqnarray}
We emphasize that the radiation boundary condition {\bf\ref{reactive}} is not 
postulated, but is rather
{\it derived} from the original 
discrete model  in the considered scaling limit. 
Using the Laplace transform method we  solved this problem exactly
and obtained the result in Eq. {\bf\ref{result2}}. In conclusion, our 
approximate
result in Eqs.{\bf\ref{result2}-\ref{coeff}} provides the {\it exact}
solution of the diffusion model \cite{millhauser,condat} in the scaling
limit! This exact analytical solution is
obtained here for the first time.
 Note, however, that this so obtained diffusion model is not able to 
resolve 
the puzzling voltage dependence in Eq. {\bf\ref{ko}}.

\section*{Synopsis and Conclusions}
With this work we put forward a unifying generalization of the
diffusion theory of ion channel gating by Millhauser {\it et al.}
\cite{millhauser,condat}.  
Our novel theory reproduces for the first time the functional
form of the puzzling voltage-dependence in Eq. {\bf\ref{ko}}.
The latter has been postulated almost fifty years ago in 
the pioneering paper by Hodgkin and Huxley \cite{HodHux52}
and is commonly used in the neurophysiology up to now.
 The proposed model  of the Fokker-Planck-Kramers type
explains the origin of
steep voltage-dependence in Eq. {\bf\ref{ko}} within a clear
physical picture which seemingly is consistent   with both our current 
understanding of the physics of proteins and  basic experimental facts.
Our study furthermore reveals the connection between the voltage dependence of 
the opening rate and the intricate behavior for the closed residence time  
distribution in corresponding voltage regimes.
A particularly appealing feature of our approach is that
our model contains only four voltage-independent physical  
parameters: the diffusion time $\tau_D$, the fraction of voltage 
-dependent substates $\xi$, the gating charge $q$ and the threshold voltage
 $V_c$. Several experimental findings could be described consistently while 
still other ones call for an experimental validation.

In particular, when (i) the activation barrier is very high,
i.e.,  $V\ll V_c$, 
the activation step  determines completely
the opening rate: the distribution of closed residence times is nearly 
exponential, as well as the voltage-dependence of the opening rate.
The channel is then predominantly closed. We remark that the opening rate should
 exhibit an exponential dependence on temperature as well. This conclusion 
follows from Eqs. {\bf\ref{result7}-\ref{bc}} and the fact that
in accord with our model the parameter $a_c$ in Eq. {\bf\ref{ko}}
is  temperature independent. Indeed,
with the diffusion time $\tau_D$ being inversely proportional
to the temperature, i.e. $\tau_D\sim 1/D
\sim 1/(k_BT)$,  yielding $a_c\sim 1/(\tau_Dk_BT)$, cf. Eq. {\bf\ref{ac}}.
In contrast, when (ii) the activation 
barrier vanishes, i.e. the voltage shifts towards the positive direction, 
the closed residence time distribution becomes non-exponential. 
On the intermediate time scale given in Eq. {\bf\ref{inter}}, 
this distribution exhibits a {\it power law} behavior, 
$f_c(t)\propto t^{-3/2}$, which crosses over 
into an exponential one at $t>\tau_D$. The emergence of the
exponential tail can be used to determine the conformational 
diffusion time $\tau_D$ experimentally.  When (iii) the
activation barrier assumes negative values at voltages $V>V_c$, 
our  result for the opening rate exhibits a 
linear dependence on voltage and, consequently, see Eq. 
{\bf\ref{result7}}, it {\it no longer} 
depends on temperature. 
The weak temperature dependence will 
 emerge however when we
renormalize the diffusion coefficient $D$ due to the roughness
of random energy landscape (cf. Fig. 2). Assuming uncorrelated 
Gaussian disorder one
gets $D\sim k_BT\exp(-\langle\delta U^2\rangle/(k_BT)^2)$
\cite{deGennes,HTB90}, where 
$\langle \delta U^2\rangle$ is the mean-squared height of the barrier
between substates. Then, $k_o \sim \exp(-\langle
\delta U^2\rangle/(k_BT)^2)$, and since $
\sqrt{\langle\delta U^2\rangle}\sim k_BT$ this non-Arrhenius
dependence is weak at room temperatures. 
This result has a clear thermodynamic
interpretation: when the activation barrier vanishes the
closed-to-open transition is entropy dominated and thus
the opening rate will only weakly depend on 
temperature. In accord with our
model this type of behavior correlates with a non-exponential 
kinetics.

The temperature behavior of the opening rate (or, equivalently, the
mean closed time) presents a true benchmark result of our theory.
The authors are looking forward to seeing this feature being tested
experimentally.

{\bf Acknowledgement}. The authors thank Peter Reimann
for fruitful discussions. 
This work has been supported by the Deutsche
Forschungsgemeinschaft via SFB 486 (Project A10).

\end{document}